\pdfoutput=1

\documentclass[11pt,twoside,a4paper,cmspaper,final,collab]{cms-tdr}
\def\svnVersion{347625}\def\cmsCernNoTag{CERN-EP/2016-068}\def\cmsCernDate{\today}\def\cmsMessage{Published in the European Physical Journal C as \href{http://dx.doi.org/10.1140/epjc/s10052-016-4149-y}{\doi{10.1140/epjc/s10052-016-4149-y.}}}

\begin{document}\cmsNoteHeader{EXO-13-002}

\hyphenation{had-ron-i-za-tion}
\hyphenation{cal-or-i-me-ter}
\hyphenation{de-vices}
\RCS$Revision: 342897 $
\RCS$HeadURL: svn+ssh://svn.cern.ch/reps/tdr2/papers/EXO-13-002/trunk/EXO-13-002.tex $
\RCS$Id: EXO-13-002.tex 342897 2016-05-13 20:19:24Z alverson $
\newlength\cmsFigWidth
\ifthenelse{\boolean{cms@external}}{\setlength\cmsFigWidth{0.85\columnwidth}}{\setlength\cmsFigWidth{0.4\textwidth}}
\ifthenelse{\boolean{cms@external}}{\providecommand{\cmsLeft}{top}}{\providecommand{\cmsLeft}{left}}
\ifthenelse{\boolean{cms@external}}{\providecommand{\cmsRight}{bottom}}{\providecommand{\cmsRight}{right}}
\cmsNoteHeader{EXO-13-002}
\title{Search for lepton flavour violating decays of heavy resonances and quantum black holes to an $\Pe\mu$ pair in proton-proton collisions at $\sqrt{s}=8\TeV$}
\titlerunning{Search for lepton flavour violating decays of heavy resonances}

\newcommand{\emu}{\Pe{}\Pgm{}}
\newcommand{\ap}{\ensuremath{\cPgg{}^\prime}}
\newcommand{\zpap}{\zp{}\xspace}
\newcommand{\WW}{\PW{}\PW{}}
\newcommand{\WZ}{\PW{}\cPZ{}}
\newcommand{\ZZ}{\cPZ{}\cPZ{}}
\newcommand{\tW}{\cPqt{}\PW{}}
\newcommand{\Wgamma}{\PW{}\cPgg{}}
\newcommand{\tautau}{\Pgt{}\Pgt{}}
\newcommand{\TSneut}{\ensuremath{\tilde{\nu}_{\Pgt}}\xspace}
\newcommand{\MTSneut}{\ensuremath{M_{\tilde{\nu}_{\Pgt}}}\xspace}
\newcommand{\Ll}{\ensuremath{\lambda_{132}}\xspace}
\newcommand{\LlA}{\ensuremath{\lambda_{132}}\xspace}
\newcommand{\LlB}{\ensuremath{\lambda_{231}}\xspace}
\newcommand{\Lq}{\ensuremath{\lambda'_{311}}\xspace}
\newcommand{\EMu}{\ensuremath{\Pe^{\pm}\Pgm^{\mp}}\xspace}
\newcommand{\etcutee}{35}
\newcommand{\gZp}{\ensuremath{\mbox{g}_{12}} \xspace}
\newcommand{\eZp}{\ensuremath{\mbox{e}_{12}} \xspace}
\newcommand{\kap}{\ensuremath{\kappa}\xspace}
\newcommand{\Klong}{\ensuremath{\mbox{K}^{0}_{\mathrm{L}}}\xspace} 
\date{\today}

\abstract{
A search for narrow resonances decaying to an electron and a muon is presented. The \emu{} mass spectrum is also investigated for non-resonant contributions from the production of quantum black holes (QBHs). The analysis is performed  using data corresponding to an integrated luminosity of 19.7\fbinv collected in proton-proton collisions at a centre-of-mass energy of 8\TeV with the CMS detector at the LHC.  With no evidence for physics beyond the standard model in the invariant mass spectrum of selected $\Pe\mu$ pairs, upper limits are set at 95$\%$ confidence level on the product of cross section and branching fraction for signals arising in theories with charged lepton flavour violation. In the search for narrow resonances, the resonant production of a \Pgt{} sneutrino in R-parity violating supersymmetry is considered. The \Pgt{} sneutrino is excluded for masses below 1.28\TeV for couplings $\lambda_{132}=\lambda_{231}=\lambda'_{311}=0.01$, and below 2.30\TeV for $\lambda_{132}=\lambda_{231}=0.07$ and $\lambda'_{311}=0.11$. These are the most stringent limits to date from direct searches at high-energy colliders. In addition, the resonance searches are interpreted in terms of a model with heavy partners of the $\PZ$ boson and the photon. In a framework of TeV-scale quantum gravity based on a renormalization of Newton's constant, the search for non-resonant contributions to the \emu{} mass spectrum excludes QBH production below a threshold mass $M_{\mathrm{th}}$ of 1.99\TeV. In models that invoke extra dimensions, the bounds range from 2.36\TeV for one extra dimension to 3.63\TeV for six extra dimensions. This is the first search for QBHs decaying into the \emu{} final state.
}

\hypersetup{%
pdfauthor={CMS Collaboration},%
pdftitle={Search for lepton flavour violating decays of heavy resonances and quantum black holes to an electron/muon pair in proton-proton collisions at sqrt(s) = 8 TeV},%
pdfsubject={CMS},%
pdfkeywords={CMS, physics, lepton flavor violation, quantum black holes}}

\maketitle

\section{Introduction} \label{sec:intro}

Several extensions of the standard model (SM)
predict the existence of heavy, short-lived states that decay to the \emu{} final state,
 and motivate the search for lepton flavour violating (LFV) signatures in interactions involving charged leptons.
This paper reports a search for phenomena beyond the SM in the invariant mass spectrum of \emu{}
pairs. The analysis is based on
data with an integrated
luminosity of 19.7\fbinv collected in proton-proton ($\Pp\Pp$) collisions at $\sqrt{s} = 8\TeV$ with the CMS detector
at the CERN LHC~\cite{Evans:2008zzb}. The results are interpreted in terms of three theoretically predicted objects:
a $\Pgt$ sneutrino ($\TSneut$) lightest supersymmetric particle (LSP) in R-parity violating (RPV) supersymmetry (SUSY) \cite{Barbier:2004ez},
interfering LFV \zp{} and \ap{} bosons
 \cite{Frere:2004yu}, and quantum black holes (QBHs) \cite{Meade:2007sz,Calmet:2008dg,Gingrich:2009hj}.

In RPV SUSY, lepton number can be violated at tree level in interactions between fermions and sfermions, and the \TSneut may be the LSP~\cite{Bernhardt:2008jz}.
For the resonant \TSneut signal, the following trilinear RPV part of the superpotential is considered:
 ${W_{\mathrm{RPV}}= \frac{1}{2} \lambda_{ijk}L_{i} L_{j} \bar{E}_{k} + \lambda'_{ijk} L_{i} Q_{j} \bar{D}_{k}}$\ifthenelse{\boolean{cms@external}}{, where $i$, $j$, and $k \in \{1,2,3\}$}{ ($i,j,k \in {1,2,3}$), where $i$, $j$, and $k$} are generation indices, $L$ and $Q$ are the $SU(2)_{L}$ doublet superfields of the leptons and quarks, and $\bar{E}$ and $\bar{D}$ are the
$SU(2)_{L}$ singlet superfields of the charged leptons and down-like quarks.
We assume that all RPV couplings vanish, except for \LlA, \LlB, and \Lq, and consider a SUSY mass hierarchy with a \TSneut LSP.
In this model, the \TSneut can be produced resonantly in $\Pp\Pp$ collisions via the \Lq coupling and it can decay either into an \emu{} pair
via the \LlA and \LlB couplings, or into a \cPqd{}\cPaqd{} pair via the \Lq coupling.
In this analysis we consider only the \emu{} final state and, for simplicity, we assume ${\LlA=\LlB}$.

The LFV \zpap{} signal is based on a model with two extra dimensions \cite{Frere:2003yv,Frere:2004yu}, where the three generations of the SM arise from a single generation in higher-dimensional space-time.
Flavour changing processes are introduced through the Kaluza-Klein modes of gauge fields that are not localised on a brane.
In four-dimensional space-time, an effective Lagrangian can be obtained that contains two complex vector fields \zp{} and \ap{}.
These vector fields generate transitions between the families in which the generation number changes by unity, such as the process\ifthenelse{\boolean{cms@external}}{
$${\cPqd + \cPaqs \rightarrow \zpap/\ap \rightarrow \Pe^{-} + \Pgm^{+}}$$
}
{
${\cPqd + \cPaqs \rightarrow \zpap/\ap \rightarrow \Pe^{-} + \Pgm^{+}}\ \ $
}and its charge conjugate. The structure of the terms in the Lagrangian for the production and decay of the
\zp{} and \ap{} bosons
is analogous to that describing the interactions
of the $\PZ$ boson and the photon with quarks and charged leptons, respectively. The coupling strengths
\gZp and \eZp are related to their SM counterparts through
a multiplicative coupling modifier~\kap.
For simplicity, the masses $M_{\zp{}}$ and $M_{\ap{}}$ are assumed to be equal, and the model is referred to as the
LFV $\zp{}$ model. It is characterized by the two independent parameters $M_{\zp{}}$ and \kap.

Theories that have a fundamental Planck scale of the order of a TeV~\cite{RandallSundrum,Randall:1999vf,ADD,ArkaniHamed:1998nn,XcalmetHsuReeb}
offer the possibility of producing microscopic black holes~\cite{Banks:1999gd,Dimopoulos:2001hw,Giddings:2001bu} at the LHC.
In contrast to semiclassical, thermal black holes, which would decay to high-multiplicity final states, QBHs are non-thermal objects expected to decay predominantly to pairs of particles.
We consider the production of a ${\mbox{spin-}0}$, colourless, neutral QBH in a model with lepton flavour violation,
in which the cross section for QBH production is extrapolated from semiclassical black holes and depends on the threshold mass $M_{\mathrm{th}}$ for QBH production and
the number of extra dimensions $n$. For ${n=0}$, it corresponds to a 3+1-dimensional model with low-scale quantum gravity, where a renormalization of Newton's constant leads to a Planck scale
at the TeV scale~\cite{XcalmetHsuReeb,Xcalmet2010,Xcalmet2014}; ${n=1}$ corresponds to the Randall--Sundrum~(RS) brane world model~\cite{RandallSundrum,Randall:1999vf}; and ${n>1}$ to the {Arkani-Hamed--Dimopoulos--Dvali~(ADD)} mo\-del \cite{ADD,ArkaniHamed:1998nn}. We consider flat-space black holes
(black holes that are spherical both in the brane and in the bulk dimensions) and, in the case of RS-type black holes ($n=1$), consider only the regime
in which almost flat five-dimensional space is an applicable metric. This is the case for $r_S \ll 1/(ke^{-kr_c})$, where $r_S$ is the Schwarzschild
radius, $k$ denotes the Anti-de Sitter curvature, and $r_c$ is the size of the extra dimension.
The threshold $M_{\mathrm{th}}$ is assumed to be at the Planck scale
in the definition of the Particle Data Group~\cite{Agashe:2014kda}
for ${n=0}$ and
${n>1}$, whereas for ${n=1}$ both the PDG and RS definitions \cite{Meade:2007sz} are adopted. In this model, the branching fraction of QBH decays to the
\EMu final state is 1.1\%, which is twice that of the dimuon or dielectron decay modes, making the \EMu signature the most promising leptonic decay channel.
While the resonant \TSneut and LFV \zpap{} signals result in a narrow peak in the invariant mass spectrum of the \emu{} pair, the mass distribution
of the QBH signal is characterized by an edge at the threshold for QBH production, and a monotonically decreasing tail.

Direct searches for resonances in the \emu{} invariant mass spectrum with interpretations in terms of \TSneut production have been carried out by the CDF \cite{CDF_emu} and D0 \cite{D0_emu} collaborations at the
Fermilab Tevatron and most recently by the
ATLAS collaboration~\cite{ATLAS_emu} using $\Pp\Pp$ collision data at a centre-of-mass energy of $8\TeV$ at the LHC. For couplings $\Ll=0.07$ and $\Lq=0.11$, the most stringent of these limits stems from the search performed by the ATLAS collaboration, excluding at 95\%~confidence level (CL) a \TSneut below a mass of
$2.0\TeV$.
Low-energy muon conversion experiments~\cite{Bertl:2006up} yield strong limits as a function of the \Pgt{} sneutrino mass on the product of the two RPV couplings of
${\Ll \Lq<3.3 \times 10^{-7} \, \left(\MTSneut/1\TeV\right)^{2}}$ at 90\% CL \cite{PhysRevD.91.055018}.
In the case of the \zpap{} signal, searches for ${\Klong \rightarrow \emu}$ decays constrain the coupling modifier \kap.
For the choice $M_{\zp{}}=M_{\ap{}}$, a bound of $\kap\lesssim M_{\zp{}}/100\TeV$ is obtained at 90\% CL~\cite{Frere:2003ye,Frere:2004yu}.
There have been searches for QBHs decaying hadronically, by the CMS~\cite{CMSQBHST,Khachatryan:2015sja,Khachatryan:2110669} and ATLAS~\cite{Aad:2014aqa,ATLAS:2015nsi} collaborations,
and in the photon plus jet, lepton plus jet, dimuon, and dielectron final states, by the ATLAS collaboration \cite{Aad:2013cva,Aad:2015ywd,Aad:2013gma,Aad:2014cka}.
This is the first search for QBH decays into the \emu{} final state.

{\tolerance=1200
The search for the phenomena beyond the SM described above is carried out for invariant masses of the \emu{} pair of ${M_{\emu} \geq 200\GeV}$,
which is the relevant region in light of existing constraints from other direct searches.
Using the same event selection, the \emu{}~invariant mass spectrum is searched for two different signal shapes:
the shape associated with a narrow resonance that may be interpreted in terms of any model involving a resonance decaying promptly into an electron and a muon, and the more model-specific QBH signal shape.
With a
relative \emu{} invariant mass resolution ranging from 1.6\%
at ${M_{\emu}=200\GeV}$ to 6\% at ${M_{\emu}=3\TeV}$, the CMS detector is a powerful tool for searches for new physics in the \emu{}~invariant mass spectrum.
}

\section{The CMS detector} \label{sec:cms}

The central feature of the CMS apparatus is a superconducting solenoid of 6\unit{m} internal diameter, providing a magnetic field of 3.8\unit{T}.
Within the solenoid volume are a silicon pixel and strip tracker, a lead tungstate crystal electromagnetic calorimeter (ECAL), and a brass and scintillator hadron calorimeter (HCAL), each composed of a barrel and two endcap sections.
Extensive forward calorimetry complements the coverage provided by the barrel and endcap detectors. Muons are measured in gas-ionization detectors embedded in the steel flux-return yoke outside the solenoid.
The silicon tracker consists of 1440 silicon pixel and 15\,148 silicon strip detector modules and measures charged particles within the pseudorapidity range ${\abs{\eta}< 2.5}$.
The ECAL consists of 75\,848 lead tungstate crystals and provides coverage for ${\abs{ \eta }< 1.479}$ in a barrel region and ${1.479 <\abs{ \eta } < 3.0}$ in two endcap regions.
Muons are measured in the range ${\abs{\eta}< 2.4}$, with detection planes using three technologies: drift tubes, cathode strip chambers, and resistive plate chambers.
A two-level trigger system is used by the CMS experiment.
The first level is composed of custom hardware processors and uses information from the calorimeters and muon detectors to select interesting events and to reduce the event rate from
the initial bunch crossing frequency of 20\unit{MHz} to a maximum of 100\unit{kHz}.
The high-level trigger processor farm further decreases the event rate to 400\unit{Hz} before data storage.
A detailed description of the CMS detector, together with a definition of the coordinate system used and the relevant kinematic variables, can be found in Ref.~\cite{Chatrchyan:2008zzk}.

\section{Event selection}
\label{sec:selection}
The search is designed in a model-independent way by requiring only one prompt, isolated muon and one prompt, isolated electron in the event selection. This minimal selection allows for
a reinterpretation of the results in terms of models with more complex event topologies than the single \emu{} pair present in the signals considered in this paper.

The data sample is selected using a single-muon trigger with a minimum transverse momentum (\pt) requirement of ${\pt > 40\GeV{}}$. In order
to allow the trigger to remain unprescaled, the pseudorapidity of the muons is constrained
to values ${|\eta|<2.1}$.
Offline, each event is required to have a
reconstructed $\Pp\Pp$ collision vertex with at least four associated tracks, located
less than 2\cm from the centre of the detector in the plane
transverse to the beam and less than 24\cm from it in the direction along the
beam. The primary vertex is defined as the vertex with the largest sum of squared transverse momenta of its associated tracks.

The reconstruction and identification of electrons and muons is carried out using standard CMS algorithms,
described in more detail
in Refs.~\cite{MUO-10-004-PAS,Khachatryan:2015hwa,EWK-10-002-PAS,CMS-dilep-2011,Khachatryan:2014fba}.
Reconstruction of the muon track starts from two tracks, one built in the silicon tracker and one built in the muon system.
Hits used to reconstruct the tracks in the two systems are then used to reconstruct a track spanning over the entire detector~\cite{MUO-10-004-PAS}.
Muon candidates are required to have a transverse momentum of ${\pt > 45 \GeV}$ with a measured
uncertainty of ${\delta(\pt)/\pt < 0.3}$ and must fall into the acceptance of the
trigger of ${|\eta|<2.1}$.
The candidate's track must have transverse and longitudinal impact parameters with respect
to the primary vertex position of less than 0.2\cm and 0.5\cm, respectively.
At least one hit in the pixel detector, six or more hits in silicon-strip
tracker layers, and matched segments in at least two muon detector planes are required to be associated with the reconstructed track.
In order to suppress backgrounds from muons within jets, the scalar $\pt$ sum of all other tracks within a
cone of size 0.3 in ${\Delta R=\sqrt{\smash[b]{(\Delta\eta)^{2}+(\Delta\phi)^{2}}}}$ (where $\phi$ is the azimuthal angle in radians)
around the muon candidate's track is required to be less than 10\% of
the candidate's $\pt$.

{\tolerance=1200
In the electron reconstruction, ECAL clusters are mat\-ched to silicon pixel detector hits, which are then used as seeds for the reconstruction of tracks in the tracker.
Electron candidates are built from clusters with associated tracks and must lie within the barrel or
endcap acceptance regions, with pseudorapidities of $|\eta|<$ 1.442 and
1.56 $<|\eta|<$ 2.5, respectively, with a transverse energy $\ET >\etcutee\GeV$.
The transverse energy is defined as the magnitude of the projection on the plane perpendicular to the beam of the electron momentum
vector normalized to the electron energy measured in the ECAL.
Misidentification of jets as electrons is suppressed by requiring that the scalar sum of the
$\pt$ of all other tracks in a cone of size 0.3 in ${\Delta R}$ around the
electron candidate's track is less than 5\GeV{}.
In addition, the sum of the $\ET$ of calorimeter energy deposits in the
same cone that are not associated with the electron candidate must be less than 3\% of the candidate's $\ET$ (plus a small $\eta$-dependent offset).
To minimise the impact of additional $\Pp\Pp$ interactions in the same bunch
crossing (pileup) on the selection efficiency, the calorimeter isolation is corrected for the average energy density in the event ~\cite{Cacciari:2007fd}.
Further reduction of electron misidentification is achieved by requiring the transverse profile of the energy deposition in the ECAL to be consistent
with the expected electron profile, and the sum of HCAL energy deposits in a cone of size 0.15 in $\Delta
R$ to be less than 5\% of the electron's ECAL energy.
The transverse impact parameter of the electron candidate's track with respect to the primary vertex must not exceed 0.02\unit{cm} and 0.05\unit{cm}, for barrel and endcap candidates,
respectively, and the track must not have more than one missing hit in the layers of the pixel detector it crossed.}

The trigger efficiency has been measured using the ``tag-and-probe''
technique in dimuon events from Z decays described in ~\cite{EWK-10-002-PAS,MUO-10-004-PAS,CMS-dilep-2011}. The trigger efficiency for muons that pass the selection requirements is
${92.9\%}$ within ${|\eta| < 0.9}$,
${83.1\%}$ within ${0.9 < |\eta| < 1.2}$, and
${80.3\%}$ within ${1.2 < |\eta| < 2.1}$.
The muon identification efficiency, including the isolation requirement, is
measured with the tag-and-probe technique applied to muons from $\Z$ boson decays using tracks in the inner silicon tracker as probes.
The same efficiency of ${95 \pm 1 \%}$ (syst) is obtained in the three pseudorapidity regions ${|\eta| < 0.9}$, ${0.9 < |\eta| < 1.2}$, and ${1.2 < |\eta| < 2.1}$,
with corresponding efficiency ratios between data and the simulation of
$0.990 \pm 0.005$ (syst),
$0.992 \pm 0.005$ (syst), and
$0.995 \pm 0.005$ (syst).
A $\pt$ range up to $300\GeV$ has been probed with the tag-and-probe method and the muon identification efficiencies
remain constant within the statistical precision, as do the corresponding efficiency ratios between data and simulation.
The evolution of the muon reconstruction and identification efficiencies
and the muon trigger efficiency for muon $\pt>300\GeV$ is based on simulation.
Using dielectron events from $\Z$ boson decays~\cite{Khachatryan:2015hwa}, the total efficiency to reconstruct
and select electrons with $\pt^{\mathrm{e}}>100\GeV$ is found to be ${88 \pm 2}$\%~(syst) in the barrel region and ${84 \pm 4\%}$~(syst) in
the endcaps. According to Monte Carlo (MC) simulation, the variation of these efficiencies with electron \pt is less than $\pm$1\% in the
barrel and $\pm$2\% in the endcaps.
The corresponding efficiency ratios for $\pt^{\mathrm{e}}>100\GeV$  between
data and simulation are $0.985 \pm 0.014$ (syst)
in the barrel and $0.981 \pm 0.004$ (syst)
in the endcaps. These efficiencies and efficiency ratios have been measured up to an electron \pt of 1\TeV in the barrel and 500\GeV in the endcap regions.

In the event selection, at least one isolated muon and one isolated electron that both pass the identification criteria
described above are required.
After the application of all efficiency scale factors that correct the simulation to the efficiencies measured in data, the combined dilepton reconstruction and identification efficiency for
RPV $\TSneut$ signal events within the detector acceptance is expected to be $80.6$\% at ${\MTSneut=200\GeV}$ and the full selection efficiency including the trigger requirement is 71.2\%.
The MC simulation predicts that this efficiency is constant within 3\% for masses between 200 \GeV and 3 \TeV.
The electron and the muon are not required to have opposite charge, in order to avoid a loss in signal efficiency due to
possible electron charge misidentification at high electron $\pt$.
Since highly energetic muons can produce bremsstrahlung resulting in an associated supercluster in the
calorimeter in the direction of the muon's inner
track, they can be misidentified as electrons.
Therefore, an electron candidate is rejected if there is a muon with $\pt$ greater
than $5\GeV$ within $\Delta R <0.1$ of the candidate.
Only one \emu{} pair per event is considered. For about $1\%$ of the events
passing the event selection there is more than one \emu{} pair in the event, in which case the pair with the highest invariant
mass is selected.

{
\section{Signal simulation} \label{sec:theory}

The RPV and QBH signal samples are generated with the \CALCHEP~(v. 3.4.1) event generator~\cite{Belyaev:2012qa}. A cross section calculation at next-to-leading order~(NLO)
 in perturbative QCD is used for the RPV signal~\cite{RPV_res_slepton_NLO}, in which
the factorization and renormalization scales are set to \MTSneut and the CTEQ6M~\cite{Pumplin:2002vw} set of parton distribution functions (PDF) is used.
The invariant mass distributions of reconstructed \emu{} pairs from simulated QBH signal samples are presented in Fig.~\ref{Fig:QBH_signal_comparison} for different signal masses and
numbers of extra dimensions. A more detailed description of the implemented QBH model including the
dependence of the $M_{\emu}$ spectrum from QBH decays on the model parameters is presented in Ref.~\cite{Belyaev_Xcalmet}.
The LFV \zpap{} signal events are produced with the \MADGRAPH~(v. 5.1.5.9) generator~\cite{Alwall:2011uj}.
The effects of the interference resulting from the $M_{\zp{}}=M_{\ap{}}$ mass degeneracy on the cross section and signal acceptance are taken into account, and
the coupling parameters of the model are taken to be the same as in Ref.~\cite{Frere:2004yu}.
All signal samples use the CTEQ6L1 \cite{Pumplin:2002vw} PDF, \PYTHIA~(v. 6.426) \cite{Sjostrand:2006za} for hadronization with the underlying event tune Z2*,
and are processed through a simulation of the full CMS detector based on \GEANTfour~(v. 9.4) \cite{Agostinelli:2002hh}.
The \PYTHIA Z2* tune is derived from the Z1 tune~\cite{Field:2010bc}, which uses the CTEQ5L PDF set, whereas Z2* adopts CTEQ6L.

The total acceptance times efficiency for each of the three signal models considered in this analysis is determined using MC simulation with selection efficiencies corrected to the values measured in data.
The signal acceptance, as defined by the selection on the lepton $\pt$ and $\eta$ applied to the generated leptons
in the signal simulation, and the product of acceptance and selection efficiency, are shown in Tables \ref{tab:accEff_numbers_resonance}
and \ref{tab:accEff_numbers_QBH}, evaluated for selected signal masses. The acceptance of the
RPV \TSneut model is that of a generic spin-0 resonance. In the case of the LFV \zpap{} model, the acceptance is more model-specific
due to the interference between the \zp{} and the \ap{}. This interference shapes the $\eta$ distributions of the leptons in the final state, which leads to a smaller acceptance
compared to a generic spin-1 resonance.
Table~\ref{tab:accEff} lists the parameterizations of the acceptance times efficiency as a function of signal mass for the RPV \TSneut and LFV \zpap{} resonance signals, resulting from fits in the mass range
from $200\GeV$ to $2.5\TeV$.
These parameterizations are used later in the statistical
interpretation of the resonance search.

\begin{figure}[htp]
\begin{center}
\includegraphics[width=.45\textwidth]{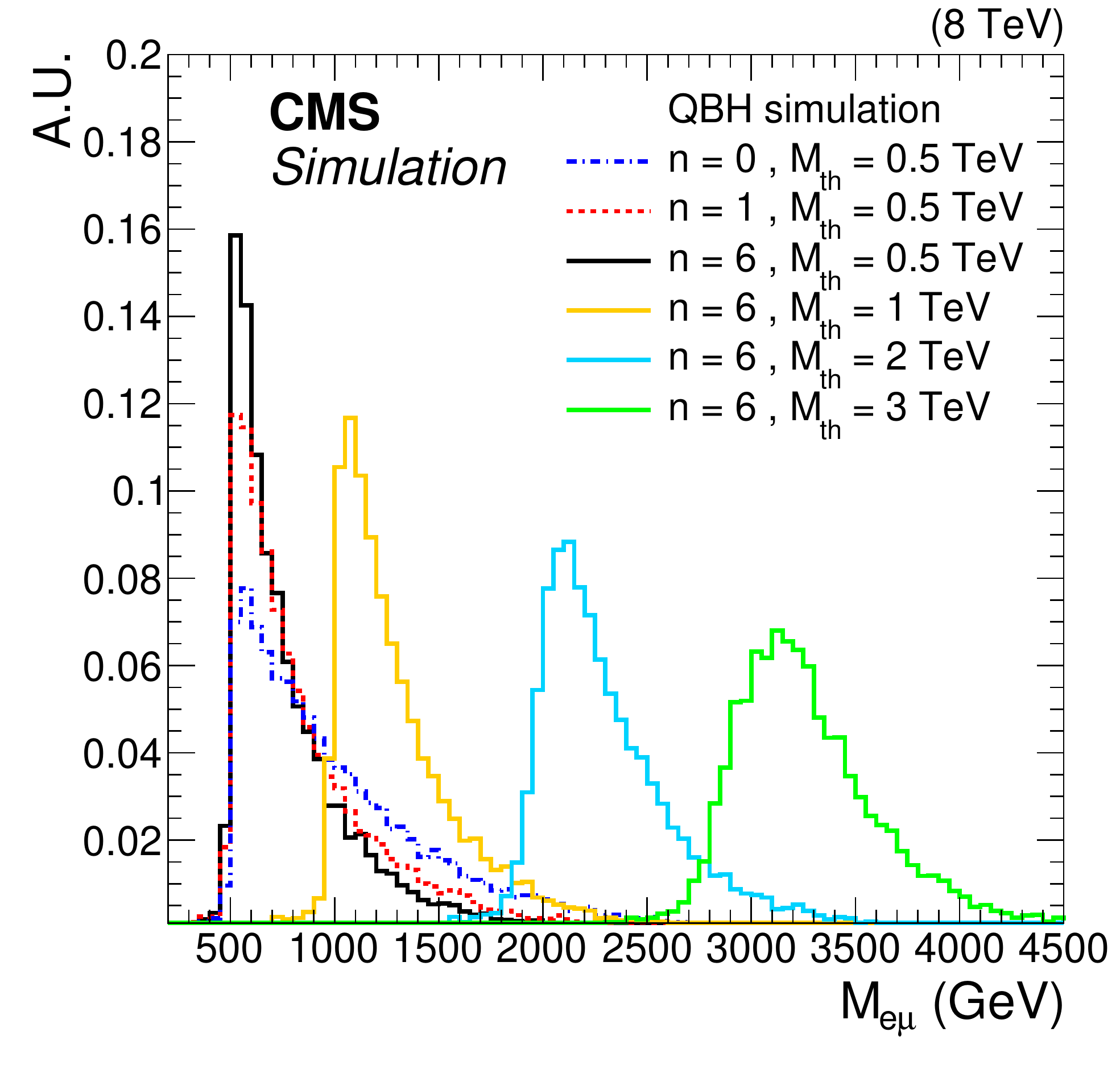}
\caption{Invariant mass distributions of reconstructed \emu{} pairs from simulated QBH signal events that pass the event selection, normalized to unit area. The steps at the threshold masses ${M_{\mathrm{th}}}$ are smeared out by the detector resolution.}
\label{Fig:QBH_signal_comparison}
\end{center}
\end{figure}

\begin{table}[htb!]
\renewcommand{\arraystretch}{1.1}
\centering
\topcaption{
Signal acceptance ($A$) and the product of acceptance and efficiency~($A \epsilon$) for different signal masses, for the RPV \TSneut and LFV \zpap{} models. The acceptance is
defined as the fraction of signal events in the simulation passing the selection on lepton $\pt$ and $\eta$ applied to the generated leptons.}
\begin{tabular}{c c c | c c c}
\hline
\MTSneut (\TeVns) & $A$ & $A \epsilon$ & $M_{\zp{}}$ (\TeVns) & $A$
& $A  \epsilon$
 \\\hline
$0.2$ & $0.59$ & $0.42$ & $0.25$ & $0.57$ & $0.39$ \\
$0.5$ & $0.80$ & $0.58$ & $0.5$ & $0.72$ & $0.51$ \\
$1.0$ & $0.89$ & $0.64$ & $1.0$ & $0.83$ & $0.59$ \\
$1.5$ & $0.91$ & $0.65$ & $1.5$ & $0.87$ & $0.61$ \\
$2.0$ & $0.92$ & $0.65$ & $2.0$ & $0.89$ & $0.62$
 \\\hline
\end{tabular}
\label{tab:accEff_numbers_resonance}
\end{table}

\begin{table}[htb!]
\renewcommand{\arraystretch}{1.1}
\centering
\topcaption{
Signal acceptance ($A$) and the product of acceptance and efficiency~($A \epsilon$) for different threshold masses~$M_{\mathrm{th}}$, for the QBH models
with $n=0$ and $n=6$ extra dimensions. The acceptance is defined as the fraction of signal events in the simulation passing the selection on lepton $\pt$ and $\eta$ applied to the generated leptons.}
\begin{tabular}{c c c | c c c}
\hline
\multicolumn{3}{c|}{ $n=0$ } & \multicolumn{3}{c}{ $n=6$ }  \\\hline
$M_{\mathrm{th}}$ (\TeVns) & $A$ & $A \epsilon$ & $M_{\mathrm{th}}$ (\TeVns) & $A$
& $A \epsilon$ \\
\hline
$0.5$ & $0.85$ & $0.61$ & $0.5$ & $0.82$ & $0.60$ \\
$1.0$ & $0.90$ & $0.63$ & $1.0$ & $0.89$ & $0.64$ \\
$2.0$ & $0.93$ & $0.64$ & $2.0$ & $0.93$ & $0.65$ \\
$3.0$ & $0.94$ & $0.63$ & $3.0$ & $0.94$ & $0.64$ \\
$4.0$ & $0.94$ & $0.62$ & $4.0$ & $0.94$ & $0.63$ \\\hline
\end{tabular}
\label{tab:accEff_numbers_QBH}
\end{table}

\begin{table}[htb!]
\renewcommand{\arraystretch}{1.1}
\centering
\topcaption{
Parametrization of the product of signal acceptance and efficiency~($A \epsilon$) as a function of signal mass $M$, for the RPV $\TSneut$ and LFV \zpap{} models. The value of $M$ is expressed in units of \GeV{}.}
\begin{tabular}{c c}
\hline
Model &Functional form of $A \epsilon$ \\
 \hline
RPV $\TSneut$ & $0.76 - 86.9/(61.4 + M) - 3.3 \times 10^{-5}~M$\\
LFV \zpap{}  & $0.74 - 141.3/(165.6 + M) - 2.7 \times 10^{-5}~M$ \\\hline	
\end{tabular}
\label{tab:accEff}
\end{table}
}

\section{Background estimation}
\label{sec:bkg}
The SM backgrounds contributing to the \emu{} final state can be divided into two classes of events. The first class comprises events with at least two prompt, isolated leptons.
The second class consists of events with either jets or photons that are misidentified as isolated leptons, and events with jets containing non-prompt leptons.
This second class of background is referred to as "non-prompt background" in this paper.
The expected SM background from processes with two prompt leptons is obtained from MC simulations.
It consists mostly of events from \ttbar{} production and \WW{} production; the former process is dominant at lower masses and the latter becomes equally
important above $M_{\emu}\sim 1\TeV$.
Other background processes estimated from MC simulation are the additional diboson processes \WZ{} and \ZZ{}, single top \tW{} production, and Drell--Yan (DY) \tautau{} events
with subsequent decay of the \tautau{} pair into an electron and a muon. The \ttbar{}, \tW{}, and \WW{} simulated samples are generated using
\POWHEG~(v. 1.0)~\cite{Nason:2004rx,Frixione:2007vw,Alioli:2010xd}
with the CT10 PDF~\cite{CT10_PDF}, and the DY, \WZ{}, and \ZZ{} background samples are generated using the {\MADGRAPH~(v.~5.1.3.30)} event generator
with the CTEQ6L1 PDF. All background samples use \PYTHIA~(v.~6.426) for hadronization with the underlying event tune $\mbox{Z2}^{*}$. The generated events are processed through a full
simulation of the CMS detector based on \GEANTfour~(v.~9.4). Pileup interactions are included in the simulation and event-dependent weights are applied in order to
reproduce the number of $\Pp\Pp$ interactions expected for the measured instantaneous luminosity. After this procedure, the distribution of the number of vertices per event observed
in data is well described by the simulation.
The simulated samples are normalized to the integrated luminosity of the data sample, $19.7~\mbox{fb}^{-1}$.
The cross sections are calculated to next-to-next-to-leading order~(NNLO) accuracy in perturbative QCD for \ttbar{} \cite{ttbar_NNLO_xsec} and DY \cite{Li:2012wna} and to NLO
accuracy for the \tW{}~\cite{Kidonakis_single_top_app_NNLO}, \WW{}, \WZ{}, and \ZZ{}~\cite{Campbell:2010ff} processes.

The main sources of non-prompt background in the \emu{} selection arise from \PW{}+jet and \Wgamma{} production with a jet or photon that are misidentified as an electron.
The \Z{}+jet, QCD multijet, and \ttbar{} processes yield subleading contributions to the background with non-prompt leptons.
The $\PW\gamma$ background is estimated from simulation based on the {\MADGRAPH~(v.~5.1.3.30)} event generator.
A background estimation based on control samples in data, using the jet-to-electron misidentification rate (MR)
method explained below, is used to determine
the $M_{\emu}$ distributions from \PW{}+jet and QCD multijet production.
The measurement of the jet-to-electron misidentification rate has been carried out in the context of Ref.~\cite{Khachatryan:2014fba}.
It starts from a sample collected using a prescaled
 single electromagnetic cluster trigger, in which the presence of an electron
candidate with relaxed electron identification criteria is required.
The events of the sample must have no more than one reconstructed electron with ${E_{\mathrm{T}} > 10\GeV}$, in order to suppress the contribution from \Z{} decays.
The misidentification measurement can be biased by selecting genuine electrons from \PW{}+jet events or converted photons
from \Pgg{}+jet events.
Processes that can give a single electron, such as \ttbar, \tW{}, \WW{}, \WZ{}, $\Z\rightarrow \tautau$,
and $\Z\rightarrow \Pe\Pe$\, where, if a second electron is produced, it
fails to be reconstructed, give another less significant source of contamination.
Simulated samples are used to correct for this contamination and its effect on the MR.
After these corrections, the electron MR, measured in bins of $E_{\mathrm{T}}$ and $\eta$, is the number of electrons passing the full selection over the
number of electron candidates in the sample.

Using the measured electron MR, the \PW{}+jet and QCD multijet contributions can be estimated from a
sample with a muon passing the single-muon trigger and the full muon selection, and an electron candidate satisfying the relaxed selection requirements but failing the full electron selection.
Each event in the sample is weighted by the factor $\mathrm{MR}/(1-\mathrm{MR})$
to determine the overall contribution of the jet backgrounds.
Contributions from processes other than \PW{}+jet and QCD multijet are subtracted from the sample to which the MR is applied, to avoid double counting.
This subtraction is based on MC simulated background samples.
A systematic uncertainty of 30\% is applied to the jet background estimate, based on cross-checks and closure tests.
An uncertainty of 50\% is assigned to the background estimate for the \Wgamma{} process, which is taken from simulation at leading order (LO) in perturbative QCD.

{
\section{Results}\label{Sec:invariant_mass_results}
After the event selection, 28~925 events are observed in data. The \emu{} invariant mass distribution is shown in Fig.~\ref{Fig:inv_mass_all}, together with the corresponding
cumulative distribution.
A comparison of the observed and expected event yields is given in Table~\ref{Tab:N_selected_events}.
The dominant background process is \ttbar{}, which contributes 69\% of the total background yield after selection,
followed by \WW{} production, contributing 11\%.
The two selected leptons carry opposite measured electric charge in 26~840
events and carry the same charge in 2085 events.
According to the background estimation, $2100\pm360$ events with same-charge \emu{} pairs are expected, most of which stem from the \PW{}+jet process,
followed by \ttbar{} and diboson production \WZ{}/\ZZ{}.
	\begin{figure*}[htp]
	\begin{center}
	\includegraphics[width=.45\textwidth]{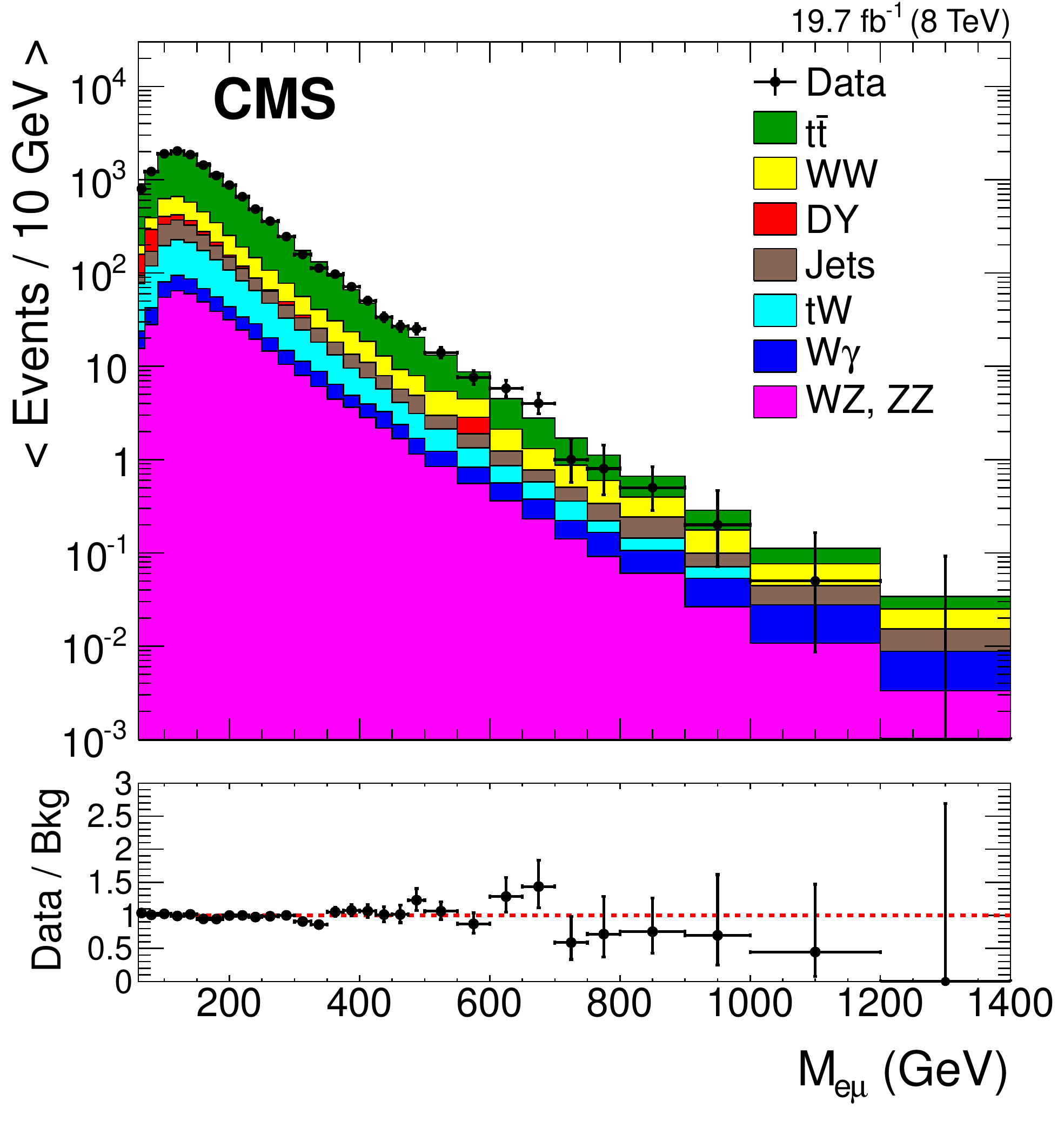}
	\includegraphics[width=.45\textwidth]{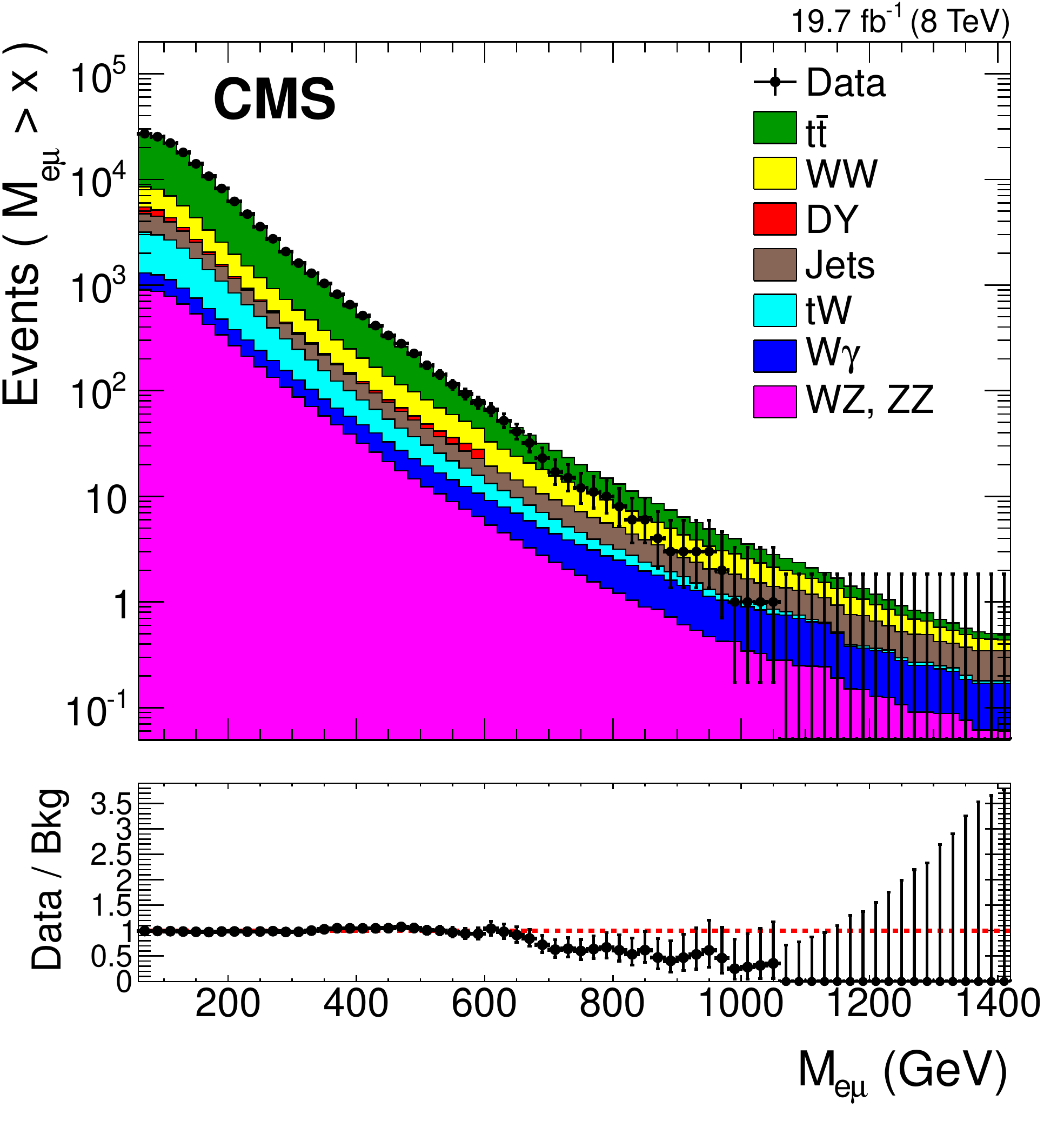}
	\caption{The invariant mass distribution of selected \emu{} pairs (left), and the corresponding cumulative distribution,
where all events above the mass value on the $x$-axis are summed (right). The points with error bars represent the data and the stacked histograms represent the expectations from SM processes.
		 The label 'Jets' refers to the estimate of the \PW{}+jet and QCD multijet backgrounds from data. The ratio of the data to the background for each bin is shown at the bottom.
The horizontal lines on the data points indicate the bin width.}
	\label{Fig:inv_mass_all}
	\end{center}
	\end{figure*}

	\begin{table*}[htp!]
	\renewcommand{\arraystretch}{1.1}
	\newcommand{\x}{\ensuremath{\phantom{0}}}
	\newcommand{\xx}{\ensuremath{\phantom{00}}}
		\topcaption{The number of observed events compared to the background expectation in five invariant mass ranges and in the full invariant mass range.
The yields obtained from simulations
are normalized according to their expected cross sections. The background label 'Jets' refers to the estimate of the \PW{}+jet and QCD multijet backgrounds from data.}
		\label{Tab:N_selected_events}
		\begin{center}
    		\begin{tabular}{c | c | c c c c c}
    		\hline
		& \multirow{2}{*}{Total}  & \multicolumn{5}{c}{Invariant mass ranges in units of \GeV{}} \\
		&  & ${<}200$ & $200$--$400$ & $400$--$600$ & $600$--$1000$ & ${>}1000$ \\ \hline
		\ttbar{} & $ 20100 \pm 1800\x $ & $15800 \pm 1400\x$ & $ 4050\pm 450\x$ & $260 \pm 44\x$ & $ 30 \pm 7\x $ & $0.9 \pm 0.4$ \\
		\WW{} & $ 3150 \pm 260\x $ & $2400 \pm 200\x$ & $670 \pm 64\x$ & $68 \pm 8\x$ & $ 13 \pm 2\x $ & $0.9 \pm 0.2$ \\
		\tW{} & $ 2000 \pm 160\x $ & $ 1550 \pm 120\x $ & $ 430 \pm 40\x $ & $ 30 \pm 3\x $ & $ \phantom{0.}4 \pm 0.5 $ & ${<}0.2$ \\
		$\mbox{Jets}$ & $ 1570 \pm 470\x $ & $1250 \pm 400\x$ & $280 \pm 83\x$ & $30 \pm 9\x$ & $ 5 \pm 2 $ & $0.6 \pm 0.3$ \\
		DY & $ 960 \pm 100 $ & $910 \pm 100$ & $40 \pm 15$ & $5 \pm 5$ & ${<}1$ & ${<}0.1$ \\
		\WZ{}/\ZZ{} & $ 940 \pm 80\x $ & $ 670 \pm 60\x $ & $ 240 \pm 20\x $ & $ 27 \pm 3\x $ & $ \phantom{0.}5 \pm 0.6 $ & $ 0.3 \pm 0.1$ \\
		\Wgamma{} & $ 480 \pm 240 $ & $360 \pm 180$ & $100 \pm 50\x$ & $12 \pm 6\x$ & $ \phantom{0.}3 \pm 1.5 $ & $0.6 \pm 0.3$ \\ \hline
		Total bkg & $29200 \pm 2300\x$ & $22900 \pm 1800\x$ & $5800 \pm 560\x$ & $430 \pm 53\x$ & $60 \pm 9\x$ & $3.5 \pm 0.6$ \\ \hline
		Data & $28925$ & $22736$ & $5675$ & $448$ & $65$ & $1$ \\ \hline
    		\end{tabular}
		\end{center}
	\end{table*}
The systematic uncertainties assigned to backgrounds obtained from simulation include the integrated luminosity (2.6\%)~\cite{CMS-PAS-LUM-13-001} and the acceptance times
efficiency (5\%).
The latter is based on the uncertainties in the various efficiency scale factors that correct the simulation to the efficiencies measured in data.
According to simulation, the evolution of the lepton selection efficiencies from the \Z{} pole,
where they are measured, to high lepton $\pt$ is covered within this uncertainty.
The uncertainty in the muon momentum scale is 5\% per \TeV{}.
Electron energy scale uncertainties are 0.6\% in the barrel and 1.5\% in the endcap.
These momentum and energy scale uncertainties cumulatively lead to an uncertainty in the total background yield of 2\% at $M_{\emu}=500\GeV$ and 3.5\% at $M_{\emu}=1\TeV$.
Uncertainties in the electron \ET and muon \pt resolutions have a negligible impact on the total background yield.
The uncertainty associated with the choice of PDF in the background simulation is evaluated according
to the PDF4LHC prescription~\cite{PDF4LHC,Botje:2011sn} and translates into an uncertainty in the background yield ranging from 5\% at $M_{\emu}=200\GeV$ to 9\% at $M_{\emu}=1 \TeV$.
Among the uncertainties in the cross sections used for the normalization of the various simulated background samples, the 5\% uncertainty in the NNLO QCD cross section of the
dominant \ttbar{} background~\cite{ttbar_NNLO_xsec} is the most relevant.
Further uncertainties associated with the modelling of the shape of the $\emu$ invariant mass distribution are taken into account for the two leading backgrounds: \ttbar{}
(higher-order corrections on the top-$\pt$ description discussed in~\cite{Kidonakis:2014pja})
and \WW{} (scale uncertainties studied with the \POWHEG generator). These lead to an uncertainty in the total background yield of up to 13\% at $M_{\emu}=1\TeV$.
A further systematic uncertainty arises from the limited sizes of the simulated background samples at high invariant mass,
where the background expectation is small.
Taking all systematic uncertainties into account, the resulting uncertainty in the background yield ranges from 9\% at $M_{\emu}=200\GeV$ to 18\% at $M_{\emu}=1\TeV$.

As shown in the cumulative invariant mass distribution in Fig.~\ref{Fig:inv_mass_all}, we observe a deficit in data compared to the background expectation for $M_{\emu}\geq700\GeV$.
In this invariant mass region, 17~events are observed and the background estimate yields $27\pm4$~(syst) events. Combining the systematic and statistical uncertainties, the local significance of
this discrepancy is below 2$\sigma$.
}

{
No significant excess with respect to the expectation is found in the measured \emu{} invariant mass distribution,
and we set limits on the product of signal cross section and branching fraction for signal mass hypotheses above $200\GeV$. Two types of signal shapes are considered for the limit setting:
a narrow resonance and the broader \emu{} invariant mass spectrum from QBH decays.
The RPV \TSneut and \zpap{} signals both result in a narrow resonance. For coupling values not excluded by existing searches,
the intrinsic widths of these signals are small compared to the detector resolution. Therefore, Gaussian functions are used to model the signal shapes.
For each probed resonance
signal mass, the two parameters, acceptance times efficiency (Table \ref{tab:accEff}) and invariant mass resolution,
define the signal shape used for limit setting.
The invariant mass resolution is derived from fits of Gaussian distributions to the \emu{}~invariant mass spectra from MC simulated signal samples
and ranges from 1.6\% at a resonance mass of ${M_{\text{res}}=200\GeV}$
to 6\% at ${M_{\text{res}}=3\TeV}$.
For high values of \emu{} pair invariant mass, it is dominated by the resolution on the measurement of the muon $\pt$,
which ranges from about 2\% at ${\pt=200\GeV}$ to 6\% at ${\pt=500\GeV}$ and
10\% at ${\pt=1\TeV}$. These values are obtained from MC simulations and agree within the uncertainties with measurements
using cosmic ray muons.
This model of the narrow resonance allows for a scan of the invariant mass spectrum with a fine spacing of the
signal mass hypothesis that corresponds to the invariant mass resolution.

Unlike the \TSneut and \zpap{} signals, the QBH signal
exhibits a broader shape with a sharp edge at the threshold mass $M_{\mathrm{th}}$ and a tail towards higher masses
(Fig. \ref{Fig:QBH_signal_comparison}).
The QBH signal shapes are obtained directly from simulated samples.

The systematic uncertainties in the signal entering the limit calculation
are the 2.6\% uncertainty in the integrated luminosity, the 5\%
uncertainty in the product of acceptance and efficiency,
and the relative uncertainty in the mass resolution, which ranges from 2\% at ${M_{\text{res}}=200\GeV}$ to 40\% at ${M_{\text{res}}=3\TeV}$.
The uncertainty in the signal acceptance times efficiency is dominated by the uncertainty in the trigger, lepton reconstruction, and identification efficiencies, and includes the
subleading PDF uncertainty in the signal acceptance.

Upper limits at $95\%$ CL on the product of cross section and branching fraction are determined using a binned likelihood Bayesian approach with a positive, uniform prior for the signal cross section \cite{ATLAS:2011tau}.
The signal and background shapes enter the likelihood with a binning of 1\GeV, well below the invariant mass resolution for masses above 200 GeV.
For the resonant signals \TSneut and \zpap{}, search regions in the invariant mass spectrum are defined as ${\pm} 6$ times the invariant mass resolution evaluated at the hypothetical resonance mass.
Only events in these search regions enter the binned likelihood in the limit calculation.
The impact of a further broadening of the signal window size on the median expected limit has been found to be negligible within the uncertainties.
For mass hypotheses above 800\GeV, the upper bound of the search region is dropped.
In the case of the QBH signal,
the search region is defined by a lower bound at ${M_{\mathrm{th}}-6\sigma_M}$, where $\sigma_M$ is the invariant mass resolution, and there is no upper bound.
The nuisance parameters associated with the systematic uncertainties are modelled with log-normal distributions, and a Markov Chain MC method is used for integration.
For each mass hypothesis considered, the posterior probability density function is derived as a function of the signal cross section
times branching fraction and yields the 95\% CL upper limit on this parameter of interest.

The $95\%$ CL limits on the signal cross section times branching fraction for the RPV \TSneut resonance signal are shown in Fig. \ref{Fig:limit_RPV_resonance_cross_section} (left).
The signal cross section shown is calculated at NLO in perturbative QCD with the RPV couplings set to $\LlA=\LlB=0.01$ and $\Lq=0.01$. For these couplings,
a lower mass limit of 1.28\TeV is obtained. At this mass, the observed limit on the cross section times branching fraction is 0.25\unit{fb}.
For a comparison with earlier searches at hadron colliders \cite{CDF_emu,ATLAS_emu},
the two coupling benchmarks $\LlA=\LlB=0.07$, $\Lq=0.11$ and $\LlA=\LlB=0.05$, $\Lq=0.10$ are considered.
For RPV couplings $\LlA=\LlB=0.07$ and $\Lq=0.11$, we set a mass limit of 2.30$\TeV$,
and improve the lower bound of 2.0$\TeV$ previously set~\cite{ATLAS_emu}.
The lower bound on the signal mass for $\LlA=\LlB=0.05$ and $\Lq=0.10$ is 2.16$\TeV$.
In the narrow width approximation, the cross section times branching fraction scales with the RPV couplings as:
\begin{equation}
\ifthenelse{\boolean{cms@external}}{
\begin{split}
\sigma \mathcal{B} \sim & \\ \left(\Lq\right)^2 &[\left(\LlA\right)^2+\left(\LlB\right)^2]/(3\left(\Lq\right)^2+[\left(\LlA\right)^2+\left(\LlB\right)^2]). \nonumber
\end{split}
 }
 {
\sigma \mathcal{B}~\sim\left(\Lq\right)^2[\left(\LlA\right)^2+\left(\LlB\right)^2]/(3\left(\Lq\right)^2+[\left(\LlA\right)^2+\left(\LlB\right)^2]). \nonumber
 }
\end{equation}

Using this relation and the observed upper cross section bounds,
we derive the limit contour in
the ${(\MTSneut,~\Lq)}$ parameter plane as a function of a fixed value of $\LlA=\LlB$.
\begin{figure*}[htp]
  \includegraphics[width=.45\textwidth]{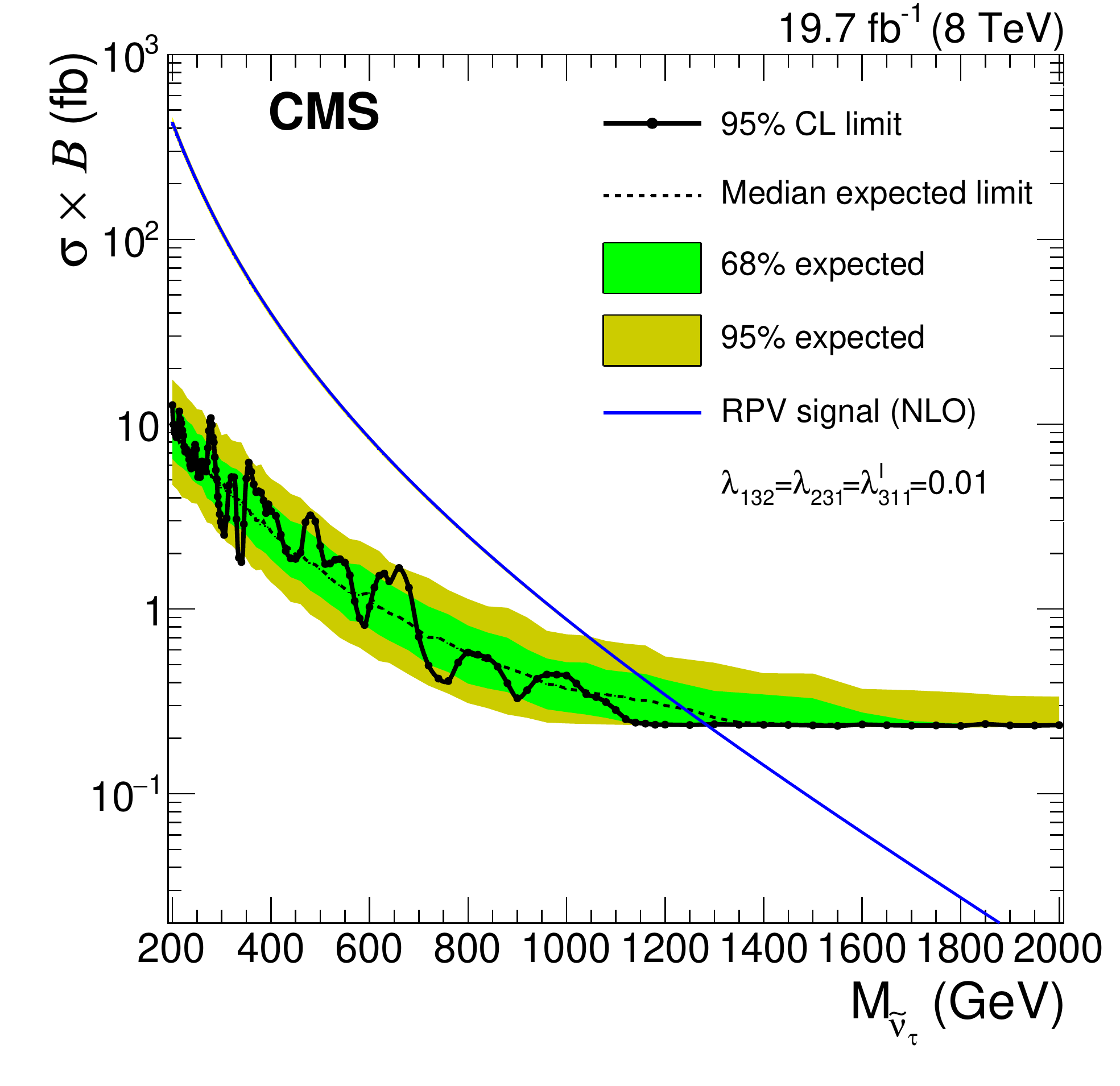}
  \includegraphics[width=.45\textwidth]{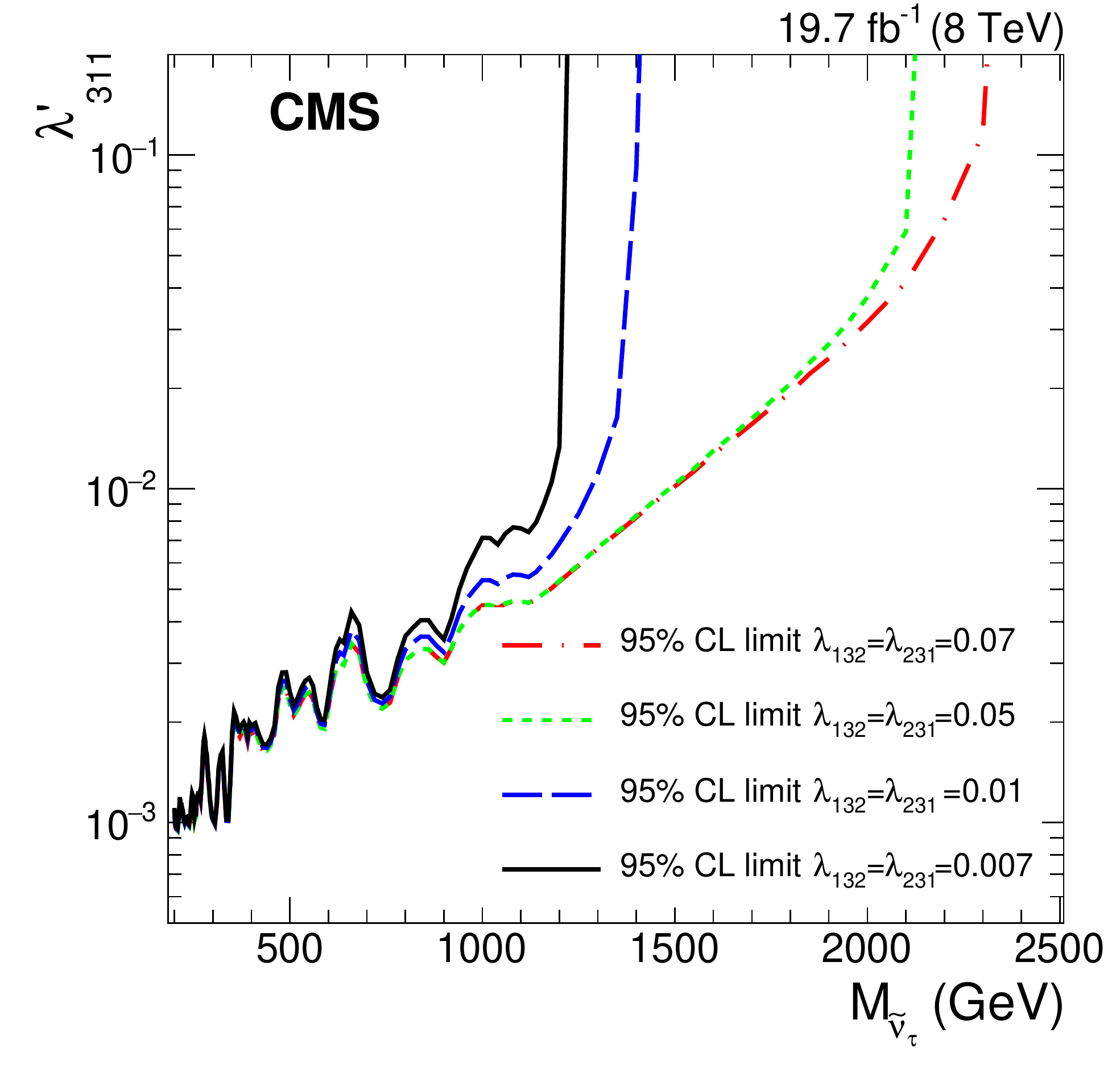}
\caption{Left: The $95\%$ CL upper limit on the product of signal cross section and branching fraction for the RPV \TSneut signal as a function of the mass of the resonance \MTSneut.
	 Right: The 95\% CL limit contours for the RPV \TSneut signal in the (\MTSneut,~\Lq) parameter plane. The values of the parameter
$\LlA=\LlB$ are fixed to $0.07$ (red dashed and dotted), $0.05$ (green small-dashed), $0.01$ (blue dashed), and $0.007$ (black solid). The regions above the curves are excluded.
}
\label{Fig:limit_RPV_resonance_cross_section}
\end{figure*}
\begin{figure*}[htp]
\includegraphics[width=.45\textwidth]{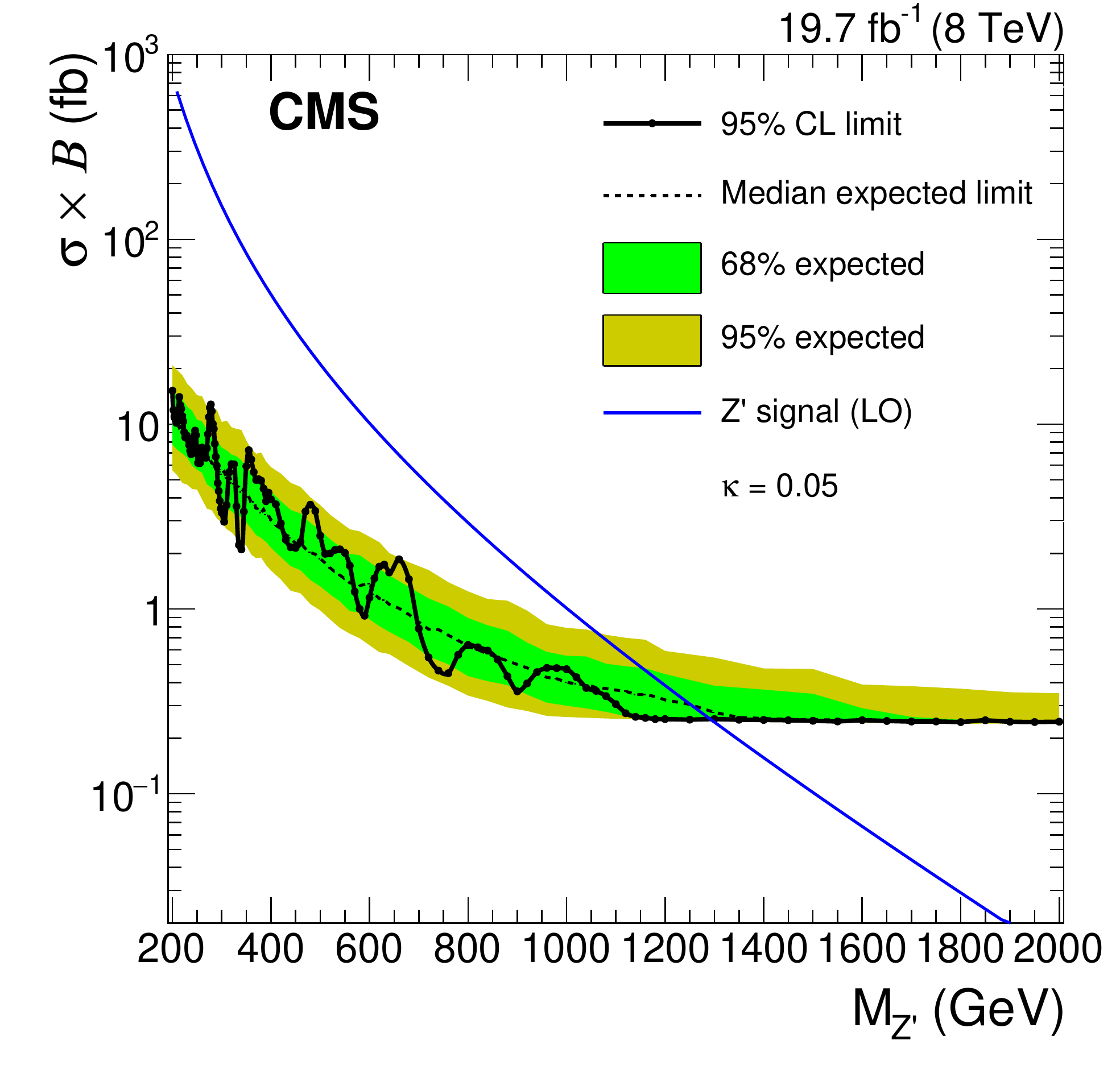}
\includegraphics[width=.45\textwidth]{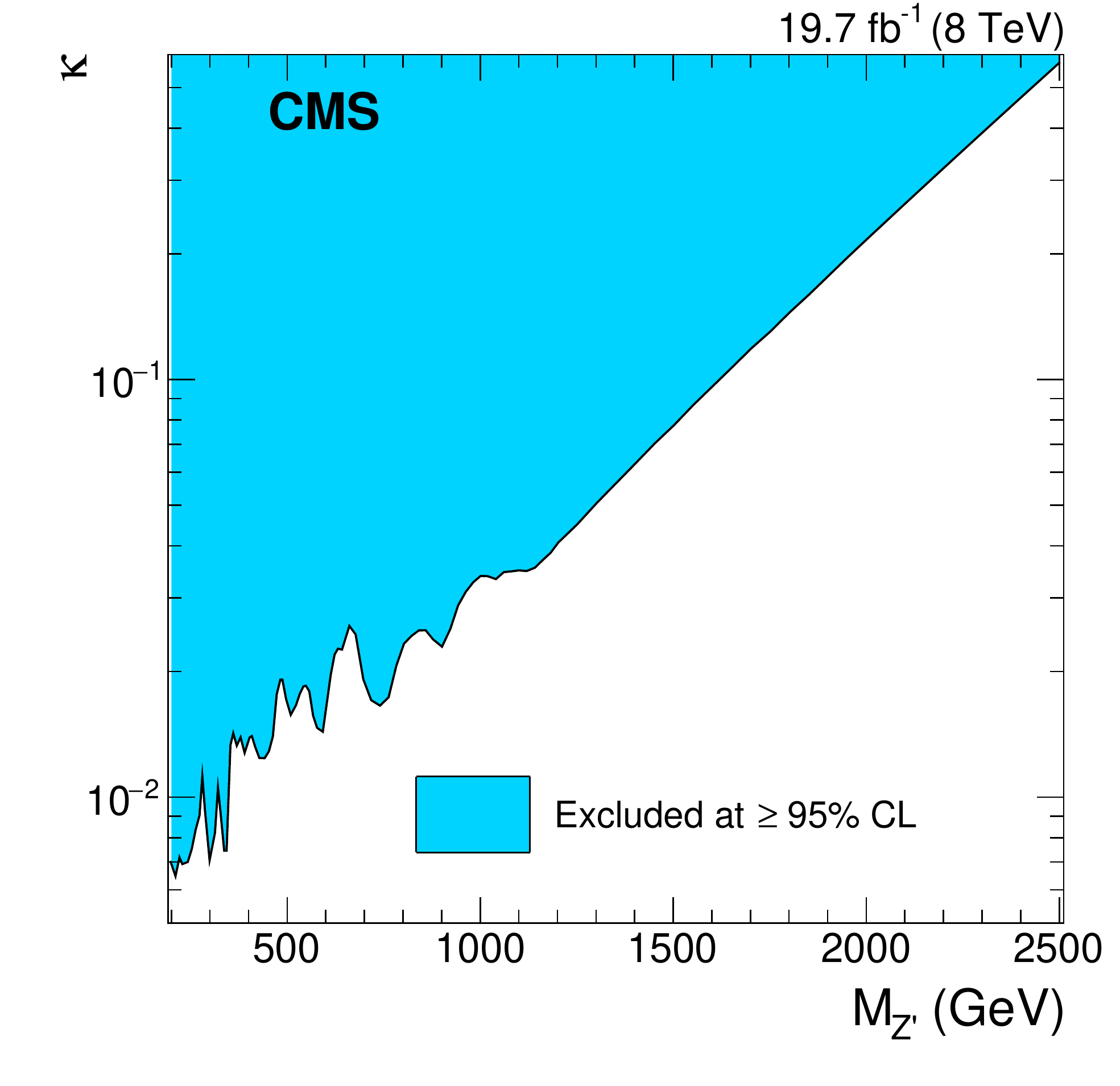}
\caption{Left: The $95\%$ CL exclusion limit on the product of signal cross section and branching fraction for the \zpap{} signal as a function of the mass $M_{\zp}$.
         Right: The 95\% CL limit contour for the \zpap{} signal in the ($M_{\zp}$,~\kap) parameter plane.}
\label{Fig:limit_zprime}
\end{figure*}
\begin{figure*}[htp]
\begin{center}
\includegraphics[width=.5\textwidth]{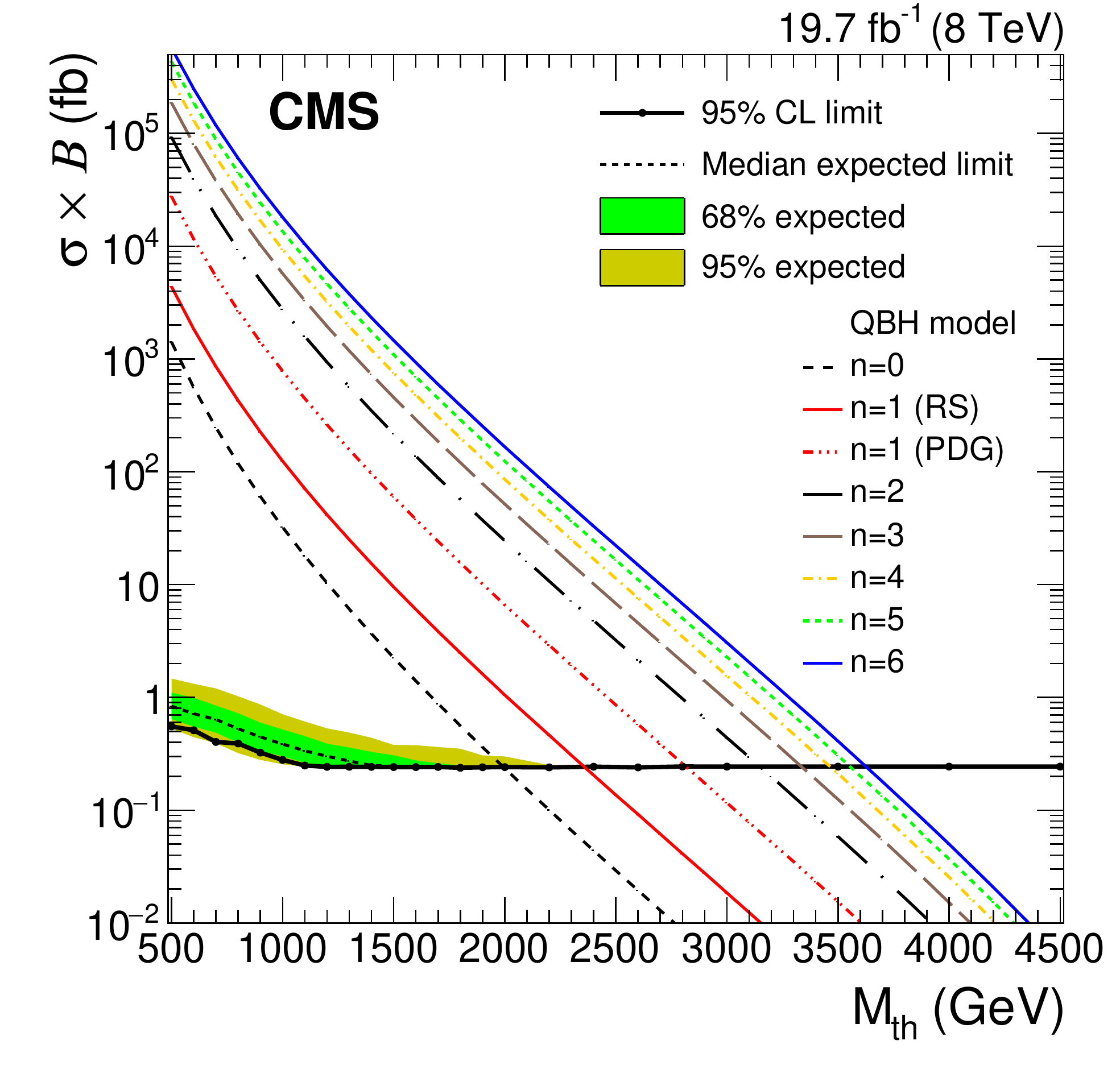}
\caption{The $95\%$ CL exclusion limit on the product of signal cross section and branching fraction for the QBH signal as a function of the threshold mass $M_{\mathrm{th}}$. The limits have been calculated using the
signal shape of the QBH model without extra dimensions ($n=0$). For signal masses ${M_{\mathrm{th}}\geq 1\TeV}$, the change in the QBH signal shape for different
numbers of extra dimensions has a negligible impact on the limit.}
\label{Fig:limit_QBH}
\end{center}
\end{figure*}
For the results presented in Fig. \ref{Fig:limit_RPV_resonance_cross_section} (right), values of the couplings \Lq and ${\LlA=\LlB}$ up to 0.2 and 0.07 are considered, respectively.
The ratio of decay width to mass of the \Pgt{} sneutrino is less than 0.5\% for these coupling values and finite-width effects are small. Searches for resonant
dijet production~\cite{Khachatryan:2015sja,Aad:2014aqa} that cover the \Pgt{} sneutrino decay to a \cPqd{}\cPaqd{} pair via the coupling \Lq do not
exclude this region of parameter space. In the model considered here with resonant production of the
\TSneut, we do not reach the sensitivity of muon conversion experiments, which lead to a bound on the coupling product of
${\Ll \Lq<3.3 \times 10^{-7} (\MTSneut/1\TeV )^{2}}$
at 90\% CL, assuming ${\LlA=\LlB}$. For comparison,
with a signal mass of ${\MTSneut=1\TeV}$ and the assumption
${\LlA=\LlB=\Lq}$, we obtain a limit of
${\Ll \Lq<4.1 \times 10^{-5}}$
at 90\% CL.
We present results in terms of the product of the production cross section and branching fraction of the \TSneut that do not depend on a specific production mechanism of the sneutrino.

The $95\%$ CL limits on the signal cross section times branching fraction for the \zpap{} signal, which exhibits a different acceptance from the spin-0 resonance in the RPV model,
are presented in Fig.~\ref{Fig:limit_zprime}~(left).
For the coupling modifier $\kap=0.05$, a lower bound on the signal mass ${M_{\zp{}}=M_{\ap{}}}$ of 1.29\TeV is obtained.
Figure~\ref{Fig:limit_zprime}~(right) shows the corresponding limit contour in the ${(M_{\zp},~\kap)}$ parameter plane.
Since this resonance is produced dominantly in the ${\cPqd\cPaqs}$ initial state, the bound from searches for muon conversion is not as strong as for the RPV \TSneut signal,
but searches for ${\Klong \rightarrow \emu}$ decays yield a stringent exclusion limit
of ${\kap\lesssim M_{\zp{}}/100\TeV}$ at 90\% CL. This can be compared to our bound of ${\kap=0.031}$ at 90\% CL for ${M_{\zp{}}=M_{\ap}=1\TeV}$.

In the QBH search, we set limits on the mass threshold for QBH production, $M_{\mathrm{th}}$, in models with $n=0$ to $n=6$ extra dimensions. The $95\%$ CL limits
on the signal cross section times branching fraction for the QBH signal are shown in Fig. \ref{Fig:limit_QBH}.
For $n=0$ in a model with a Planck scale at the TeV scale from a renormalization of the gravitational constant, we exclude QBH production below a threshold mass $M_{\mathrm{th}}$ of 1.99\TeV.
For $n=1$, two signal cross sections are considered with the Schwarzschild radius evaluated in the RS and PDG conventions. The resulting limits on $M_{\mathrm{th}}$
are 2.36\TeV and 2.81\TeV, respectively.
For ADD-type black holes with $n>1$, we obtain lower bounds on $M_{\mathrm{th}}$ ranging from 3.15\TeV for $n=2$ to 3.63\TeV for $n=6$.
A summary of the 95\% CL lower mass limits set for all signal models is presented in Table~\ref{tab:massLimits}.
\begin{table*}[htb]
\renewcommand{\arraystretch}{1.1}
\centering
\topcaption{
The ${95\%}$ CL observed and expected lower bounds on the signal masses of \Pgt{} sneutrinos in RPV SUSY, resonances in the LFV \zpap model, and QBHs, each with subsequent decay into an \emu{} pair.
For the QBH signal with $n=1$, two signal cross sections are considered with
the Schwarzschild radius evaluated in either the Randall-Sundrum~(RS) or the Particle Data Group~(PDG) convention.}
\begin{tabular}{c c c}
\hline
\multirow{2}{*}{Signal model} & \multicolumn{2}{c}{ Lower limit signal mass (TeV) } \\
 & observed & expected \\\hline
RPV \TSneut  ($\LlA = \LlB = \Lq =0.01$) & 1.28 & 1.24 \\
RPV \TSneut  ($\LlA = \LlB = 0.05~,~\Lq =0.10$) & 2.16 & 2.16 \\
RPV \TSneut  ($\LlA = \LlB = 0.07~,~\Lq =0.11$) & 2.30 & 2.30  \\ \hline
LFV \zpap ($\kap=0.05$) & 1.29 & 1.25  \\ \hline
QBH \quad $n=0$ & 1.99 & 1.99 \\
\hspace{0.72cm} QBH \quad $n=1$ (RS) & 2.36 & 2.36 \\
\hspace{1.08cm} QBH \quad $n=1$ (PDG) & 2.81 & 2.81 \\
QBH \quad $n=2$ & 3.15 & 3.15 \\
QBH \quad $n=3$ & 3.34 & 3.34 \\
QBH \quad $n=4$ & 3.46 & 3.46 \\
QBH \quad $n=5$ & 3.55 & 3.55 \\
QBH \quad $n=6$ & 3.63 & 3.63  \\\hline
\end{tabular}
\label{tab:massLimits}
\end{table*}
}

\section{Summary}
{\tolerance=1200
A search has been reported for heavy states decaying prom\-ptly into an electron and a muon using 19.7\fbinv
of proton-proton collision data recorded with the CMS detector
at the LHC at a centre-of-mass energy of $8\TeV$. Agreement is observed between the data and the standard model expectation
with new limits set on resonant production of \Pgt{} sneutrinos in R-parity violating supersymmetry with subsequent decay into \emu{} pairs.
For couplings ${\LlA=\LlB=0.01}$ and ${\Lq=0.01}$, \Pgt{} sneutrino lightest supersymmetric particles for masses \MTSneut below ${1.28\TeV}$ are excluded at 95\% CL.
For couplings ${\LlA=\LlB=0.07}$ and ${\Lq = 0.11}$, masses \MTSneut below ${2.30\TeV}$ are excluded.
These are the most stringent limits from direct searches at high-energy colliders.
For the \zpap{} signal model, a lower mass limit of $M_{\zp}=M_{\ap}={1.29\TeV}$ is set at 95\% CL for the coupling modifier $\kap=0.05$.
This direct search for resonant production of an \emu{} pair at the TeV scale does not reach the sensitivity of dedicated low-energy experiments, but complements such indirect searches and
can readily be interpreted in terms of different signals of new physics involving a heavy state that decays promptly into an electron and a muon.
Lower bounds are set on the mass threshold for the production of
quantum black holes with subsequent decay into an \emu{} pair in models with zero to six extra dimensions,
assuming the threshold mass to be at the Planck scale, ranging from ${M_{\mathrm{th}}=1.99\TeV~(n=0)}$ to ${3.63\TeV~(n=6)}$.
These are the first limits on quantum black holes decaying into \emu{} final states.
}

\newpage

\begin{acknowledgments}

We congratulate our colleagues in the CERN accelerator departments for the excellent performance of the LHC and thank the technical and administrative staffs at CERN and at other CMS institutes for their contributions to the success of the CMS effort. In addition, we gratefully acknowledge the computing centres and personnel of the Worldwide LHC Computing Grid for delivering so effectively the computing infrastructure essential to our analyses. Finally, we acknowledge the enduring support for the construction and operation of the LHC and the CMS detector provided by the following funding agencies: BMWFW and FWF (Austria); FNRS and FWO (Belgium); CNPq, CAPES, FAPERJ, and FAPESP (Brazil); MES (Bulgaria); CERN; CAS, MoST, and NSFC (China); COLCIENCIAS (Colombia); MSES and CSF (Croatia); RPF (Cyprus); MoER, ERC IUT and ERDF (Estonia); Academy of Finland, MEC, and HIP (Finland); CEA and CNRS/IN2P3 (France); BMBF, DFG, and HGF (Germany); GSRT (Greece); OTKA and NIH (Hungary); DAE and DST (India); IPM (Iran); SFI (Ireland); INFN (Italy); MSIP and NRF (Republic of Korea); LAS (Lithuania); MOE and UM (Malaysia); CINVESTAV, CONACYT, SEP, and UASLP-FAI (Mexico); MBIE (New Zealand); PAEC (Pakistan); MSHE and NSC (Poland); FCT (Portugal); JINR (Dubna); MON, RosAtom, RAS and RFBR (Russia); MESTD (Serbia); SEIDI and CPAN (Spain); Swiss Funding Agencies (Switzerland); MST (Taipei); ThEPCenter, IPST, STAR and NSTDA (Thailand); TUBITAK and TAEK (Turkey); NASU and SFFR (Ukraine); STFC (United Kingdom); DOE and NSF (USA).

Individuals have received support from the Marie-Curie programme and the European Research Council and EPLANET (European Union); the Leventis Foundation; the A. P. Sloan Foundation; the Alexander von Humboldt Foundation; the Belgian Federal Science Policy Office; the Fonds pour la Formation \`a la Recherche dans l'Industrie et dans l'Agriculture (FRIA-Belgium); the Agentschap voor Innovatie door Wetenschap en Technologie (IWT-Belgium); the Ministry of Education, Youth and Sports (MEYS) of the Czech Republic; the Council of Science and Industrial Research, India; the HOMING PLUS programme of the Foundation for Polish Science, cofinanced from European Union, Regional Development Fund; the Mobility Plus programme of the Ministry of Science and Higher Education (Poland); the OPUS programme of the National Science Center (Poland); MIUR project 20108T4XTM (Italy); the Thalis and Aristeia programmes cofinanced by EU-ESF and the Greek NSRF; the National Priorities Research Program by Qatar National Research Fund; the Rachadapisek Sompot Fund for Postdoctoral Fellowship, Chulalongkorn University (Thailand); the Chulalongkorn Academic into Its 2nd Century Project Advancement Project (Thailand); and the Welch Foundation, contract C-1845.

\end{acknowledgments}

\newpage

\bibliography{auto_generated}

\cleardoublepage \appendix\section{The CMS Collaboration \label{app:collab}}\begin{sloppypar}\hyphenpenalty=5000\widowpenalty=500\clubpenalty=5000\textbf{Yerevan Physics Institute,  Yerevan,  Armenia}\\*[0pt]
V.~Khachatryan, A.M.~Sirunyan, A.~Tumasyan
\vskip\cmsinstskip
\textbf{Institut f\"{u}r Hochenergiephysik der OeAW,  Wien,  Austria}\\*[0pt]
W.~Adam, E.~Asilar, T.~Bergauer, J.~Brandstetter, E.~Brondolin, M.~Dragicevic, J.~Er\"{o}, M.~Flechl, M.~Friedl, R.~Fr\"{u}hwirth\cmsAuthorMark{1}, V.M.~Ghete, C.~Hartl, N.~H\"{o}rmann, J.~Hrubec, M.~Jeitler\cmsAuthorMark{1}, V.~Kn\"{u}nz, A.~K\"{o}nig, M.~Krammer\cmsAuthorMark{1}, I.~Kr\"{a}tschmer, D.~Liko, T.~Matsushita, I.~Mikulec, D.~Rabady\cmsAuthorMark{2}, N.~Rad, B.~Rahbaran, H.~Rohringer, J.~Schieck\cmsAuthorMark{1}, R.~Sch\"{o}fbeck, J.~Strauss, W.~Treberer-Treberspurg, W.~Waltenberger, C.-E.~Wulz\cmsAuthorMark{1}
% ...
\vskip\cmsinstskip
% ...
1:~~Also at Vienna University of Technology, Vienna, Austria\\
2:~~Also at CERN, European Organization for Nuclear Research, Geneva, Switzerland\\
% ...

\end{sloppypar}
\end{document}